\documentclass[twocolumn,floatfix,aps,prd,showpacs]{revtex4}

\usepackage{graphicx}
\usepackage{bm}% bold math
\usepackage{epsf}

\newcommand{\beq}{\begin{equation}}
\newcommand{\eeq}{\end{equation}}
\newcommand{\beqn}{\begin{eqnarray}}
\newcommand{\eeqn}{\end{eqnarray}}
\newcommand{\pa}{\partial}

\begin{document}

\title{High-velocity collision of two black holes}

\author{Masaru Shibata, Hirotada Okawa, and Tetsuro Yamamoto}
\affiliation{Graduate School of Arts and Sciences, University of
Tokyo, Komaba, Meguro, Tokyo 153-8902, Japan}

\begin{abstract}
We study nonaxisymmetric collision of two black holes (BHs) with a
high velocity $v=|dx^i/dx^0|=0.6$--$0.9c$ at infinity, where $x^{\mu}$
denotes four-dimensional coordinates.  We prepare two boosted BHs for
the initial condition which is different from that computed by a
simple moving-puncture approach. By extrapolation of the numerical
results, we find that the impact parameter has to be smaller that
$\approx 2.5GM_0/c^2$ for formation of a BH in the collision for $v
\rightarrow c$, where $M_0 c^2$ is the initial total ADM mass energy
of the system. For the critical value of the impact parameter,
20--30\% of mass energy and 60--70\% of angular momentum are
dissipated by gravitational radiation for $v=0.6$--$0.9c$.
\end{abstract}

\maketitle

%\section{Introduction}

\noindent
{\bf\em I Introduction}: 
Clarifying formation process of mini black hole (BH) in
higher-dimensional spacetimes has become an important issue since a
possibility of BH formation in accelerators was pointed out. If our
space is the 3-brane in large \cite{ADD98} or warped \cite{RS99} extra
dimensions, the Planck energy could be of $O({\rm TeV})$ that may be
accessible with planned particle accelerators. In the presence of the
extra dimensions, a BH of very small mass may be produced in the
accelerators and the evidence may be detected.

The possible phenomenology of a BH produced in accelerators was first
discussed in \cite{BHUA} (see \cite{reviews} for reviews).  During the
high-energy particle collision of sufficiently small impact parameter
in a higher-dimensional spacetime, two particles will merge to form a
distorted BH, and then, it settles down to a quasistationary state
after emission of gravitational waves. The quasistationary BH will be
soon evaporated by the Hawking radiation, implying that the quantum
gravity effects will be important.  The evaporation and quantum
gravity effects \cite{greybody} have been studied for yielding a
plausible scenario (cf. \cite{related} for related issues). By
contrast, the analyses for BH formation and subsequent evolution by
gravitational radiation have not yet been done in detail (but see
\cite{HEADON}).  These phases are described well in the context of
general relativity \cite{GR04}, but due to its highly nonlinear
nature, any approximation breaks down. Thus, numerical relativity
simulation is the unique approach to study this phase.

In this paper, we present a new numerical-relativity study for
high-velocity collision of two BHs in four dimensions. This is the
first step for understanding the high-velocity collision of two BHs in
higher-dimensional spacetimes. We perform simulations for two
equal-mass BHs of no spin. A new approach is adopted for preparing the
initial condition (see Sec. II).  The velocity of each BH is chosen in
the range, 0.6--$0.9c$ (where $c$ is the speed of light), and the
results are extrapolated to infer a result for $v \rightarrow c$. We
determine the largest value of the impact parameter for formation of
new BH and approximately determine the final mass and spin of the
BH. In the following, we use the geometrical units in which $c=G=1$.

%\section{Initial condition}

\noindent
{\bf\em II Initial condition}: 
There are several methods for preparing initial condition for
high-velocity collision of two BHs. A popular method is the
moving-puncture approach which has been adopted in a recent work for
the head-on collision of two high-velocity BHs \cite{HEADON}. In the
simple moving-puncture approach in which the three-spatial
hypersurface is assumed to be initially conformally flat, the BHs are
not in a stationary state in their own comoving frame and hence a
large amount of spurious gravitational waves are included, as pointed
out in \cite{HEADON}. To avoid this unsuitable property, in this
paper, we use a {\em different} approach from the simple
moving-puncture one: We superimpose two boosted BHs, as described
below.

The line element of a nonrotating BH in the isotropic coordinates 
is written as
\beqn
ds^2=-\alpha_0^2 dt_0^2 + \psi_0^4 (dx_0^2 + dy_0^2 + dz_0^2),
\eeqn
where 
\beqn
\alpha_0=\Bigl(1-{m_0 \over 2r_0}\Bigr)\psi_0^{-1},~~\psi_0 =1+{m_0 \over 2r_0}.
\eeqn
$r_0=\sqrt{x_0^2 + y_0^2 + z_0^2}$ and $m_0$ is the BH mass 
in the rest frame of the BH. By boosting the BH in the 
$x$-axis direction with the speed $v$, the line element becomes
\beqn
&&ds^2=-\Gamma^2 (\alpha_0^2-\psi_0^4v^2) dt^2 
+ 2 \Gamma^2 v (\alpha_0^2 - \psi_0^4) dt dx \nonumber \\
&&~~~~~~~~~~+ \psi_0^4 (B_0^2 dx^2 + dy^2 + dz^2), \label{lorentz0}
\eeqn
where the new coordinates $x^{\mu}$ are related to the original one 
by the Lorentz transformation as
$t = \Gamma (t_0 + v x_0)$, $x = \Gamma (x_0 + v t_0)$, $y=y_0$, 
and $z=z_0$. 
%\beqn
%&&t = \Gamma (t_0 + v x_0),\\
%&&x = \Gamma (x_0 + v t_0),\\
%&&y = y_0,\\
%&&z = z_0.
%\eeqn
$\Gamma$ is the Lorentz factor, $\Gamma=1/\sqrt{1-v^2}$, and 
\beqn
B_0^2=\Gamma^2 (1-v^2 \alpha_0^2 \psi_0^{-4}).
\eeqn 
Note that at $t=0$, $r_0=\sqrt{\Gamma^2 x^2 + y^2 + z^2}$. 

Because the exact solution of a boosted BH is described by
Eq. (\ref{lorentz0}), in the following, we consider initial data for
which the spatial metric has the form
\beqn
dl^2 = \psi^4 (B^2 dx^2  + dy^2 +dz^2). \label{line}
\eeqn
Before going ahead, we summarize the lapse function ($\alpha$), 
nonzero-component of the shift vector ($\beta^i$), and non-zero 
components of extrinsic curvature ($K_{ij}$) of the boosted BH at $t=0$: 
\beqn
&& \alpha=\alpha_0 B_0^{-1},~~
\beta^x={\alpha_0^2 -\psi_0^4 \over \psi_0^4 -\alpha_0^2 v^2}v,\\
&& K_{xx}={\Gamma^2 B_0 x v \over r_0} 
\biggl[2 \alpha_0'-{\alpha_0 \over 2}[\ln (\psi_0^4-\alpha_0^2 v^2)]' \biggr]
,\label{kxx}\\
&& K_{yy}=K_{zz}={2 \Gamma^2 x v \alpha_0 \psi_0' \over \psi_0 B_0 r_0}, \\
&& K_{xy}={B_0 v y \over r_0}\biggl[\alpha_0' 
-{\alpha_0 \over 2}[\ln (\psi_0^4-\alpha_0^2 v^2)]' \biggr],\\
&& K_{xz}={B_0 v z \over r_0}\biggl[\alpha_0' 
-{\alpha_0 \over 2}[\ln (\psi_0^4-\alpha_0^2 v^2)]' \biggr].\label{kxz}
\eeqn
The dash (${}'$) denotes the ordinary derivative with respect to 
$r_0$ (e.g., $\alpha_0'=d\alpha_0/dr_0$), and 
$K_{ij}$ is derived from 
\beqn
K_{ij}={1 \over 2 \alpha}\Bigl(D_i \beta_j + D_j \beta_i -\pa_t \gamma_{ij}
\Bigr). 
\eeqn
$D_i$ denotes the covariant derivative with respect to three 
metric, $\gamma_{ij}$,  
and we use the relation $\pa_t \gamma_{ij}=-\Gamma v \pa_{x_0} \gamma_{ij}$. 

Now, we describe initial data for two BHs.  Although we adopt initial
condition which {\em approximately} satisfies constraint equations in
general relativity in this paper, a general framework is first
summarized.

We write the conformal factor as 
\beqn
%\psi = 1 + {m_1 \over 2r_1} + {m_2 \over 2r_2} + \phi,
&&\psi  = \psi_{\rm main} + \phi, \\
&&\psi_{\rm main} \equiv 1 + {m_1 \over 2r_1} + {m_2 \over 2r_2},
\eeqn
where $m_a$ denotes mass parameter of each BH, 
$r_a = \sqrt{\Gamma^2 (x-x_a)^2 + (y-y_a)^2 + z^2}$, 
and $(x_a, y_a, 0)$ denotes the location of each BH at $t=0$. 
Namely, we express each BH by a moving-puncture framework in a 
modified form, in which $r_a \not=\sqrt{(x-x_a)^2 + (y-y_a)^2 + z^2}$. 
$\phi$ is a correction term which should be determined by solving the 
Hamiltonian constraint. 

This paper focuses on the equal-mass case in which $m_1=m_2=m_0$,
$x_2=-x_1=x_0~(\geq 0)$, $y_2=-y_1=b/2~(>0)$. Two BHs are assumed to
have the same absolute velocity but move in the opposite directions
each other; i.e., $v_1=-v_2=v~(>0)$.  Here, $b$ denotes the impact
parameter. The total mass energy $M_0$ and angular momentum $J$ of the
system are
\beqn
M_0 = 2m_0 \Gamma~~{\rm and}~~J=  m_0 \Gamma v b,
\eeqn
and the nondimensional spin parameter of the system is 
\beqn
{J \over M_0^2} ={b v \over 4m_0 \Gamma}. \label{spin}
\eeqn
It is natural to expect that a new BH is formed after the collision 
whenever $J/M_0^2 <1$, i.e., $b < 4m_0\Gamma/v$. 

Taking into account that the line element of a boosted 
BH is written by Eq.~(\ref{lorentz0}), we write $B^2$ as 
\beqn
B^2=\Gamma^2 \biggl[1-v^2 
\Big(1 - {m_1 \over 2r_1} - {m_2 \over 2r_2}\Big)^2 \psi_{\rm main}^{-6}\biggr]. 
\eeqn
For the extrinsic curvature, we basically superimpose two parts as 
\beqn
K_{ij} = K_{1ij}+K_{2ij}+\delta K_{ij},
\eeqn
where $K_{aij}~(a=1, 2)$ are defined from Eqs. (\ref{kxx})--(\ref{kxz}) 
by replacing $r_0$ to $r_a~(a=1, 2)$. $\delta K_{ij}$ is a 
correction term which should be determined by solving the momentum 
constraint. 

\begin{figure}[t]
\vspace{-12mm}
\epsfxsize=3.in
\leavevmode
\epsffile{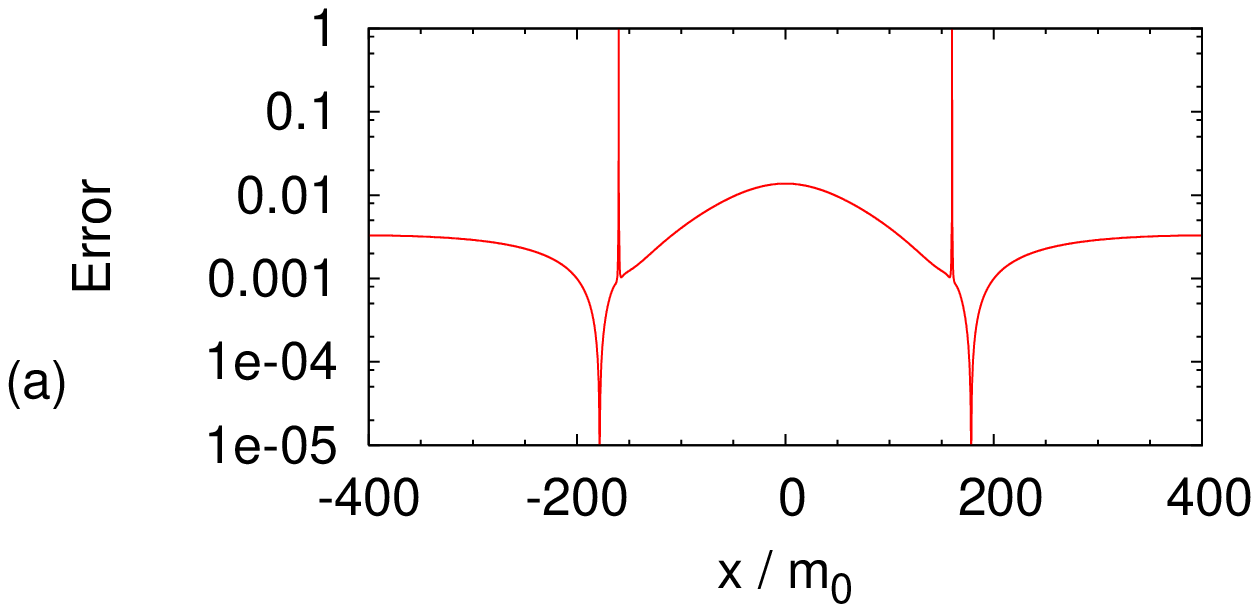} \\
\vspace{-8mm}
\epsfxsize=3.in
\leavevmode
\epsffile{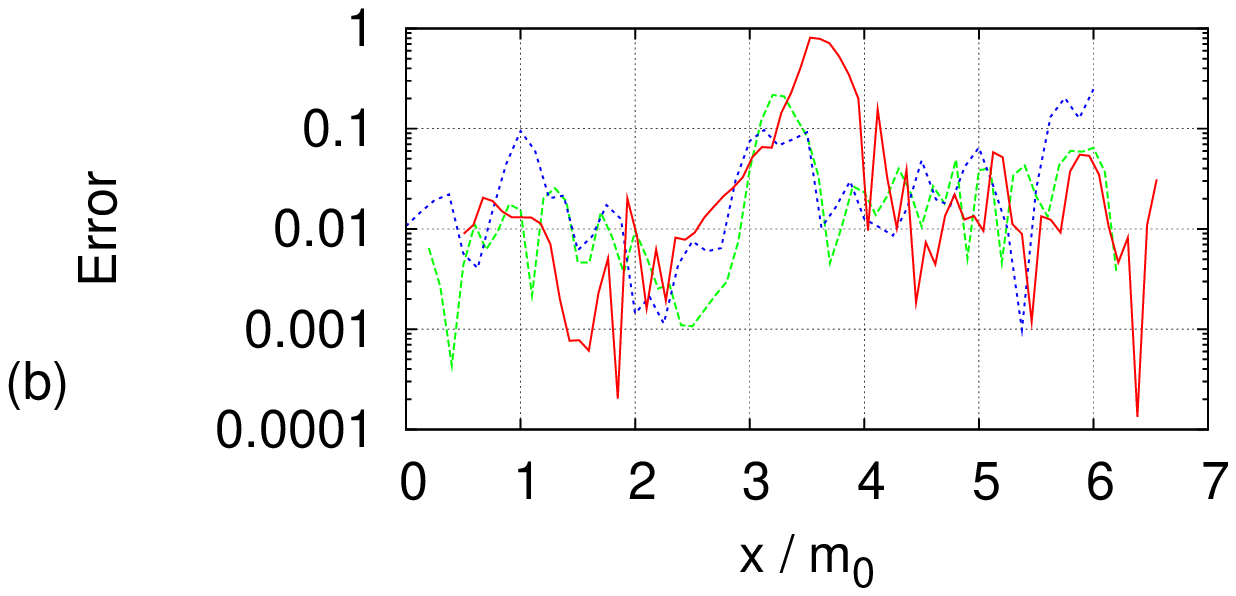} 
%\end{center}
\vspace{-12mm}
\caption{(a) Violation of the Hamiltonian constraint along $x$ axis
  for $x_0=160m_0$, $b=0$, and $v=0.9$. The violation is defined by
  $|H|/(|H_1|+|H_2|+|H_3|)$ where $H_1=\tilde D_i \tilde D^i \psi$,
  $H_2=-\tilde R \psi/8$, $H_3=[K_{ij}K^{ij}-(K_k^{~k})^2]\psi^5/8$,
  and $H=H_1+H_2+H_3$ should be zero if the Hamiltonian constraint is
  satisfied. $\tilde D_i$ and $\tilde R$ are the covariant derivative
  and Ricci scalar with respect to the conformal three metric $\tilde
  \gamma_{ij}=\psi^{-4}\gamma_{ij}$. (b) The same as (a) but for
  numerical results near one of BHs just before collision for
  $x_0=160m_0$, $bv/(m_0\Gamma)=5.088$, and $v=0.8$ in the simulations
  of different grid resolution, $h/m_0=0.0625$ (dotted curve; at
  $t=256m_0$), 0.050 (dashed curve; at $t=256m_0$), and 0.042 (solid
  curve; at $t=258m_0$). BH is located for $3 \alt x/m_0 \alt 4$. Note
  that convergence is not seen because the degree of the constraint
  violation seems to be primarily determined by the initial condition.
\label{FIG0}}
\end{figure}

In this paper, we adopt an approximate initial condition of $\phi=0$
and $\delta K_{ij}=0$.  The adopted initial data does not satisfy the
constraint equations of general relativity for finite values of $x_a$
and $y_a$ or for nonzero value of $v$. However, for the case that $R_a
\equiv (x_a^2 + y_a^2)^{1/2} /m_0 \gg 1$, the violation of the
constraints is tiny because the magnitude of the violation is
proportional to $m_0/R_a$.  In the present work, we choose $R_a/m_0
\agt 100$ (typically 160) for the initial condition.  In such case,
the violation of the constraints is $\sim m_0/R_a=O(0.01)$ for most of
region (see Fig. \ref{FIG0}), and thus, the initial condition
approximately satisfies the constraints. The exception occurs around
the puncture for which the violation is large, but the worst region is
hidden inside apparent horizon and hence does not play a bad role.
The constraint violation does not disappear during the evolution, but
the magnitude of the violation remains roughly in the same small level
at least before collision of two BHs (see Fig. \ref{FIG0}(b)).  In
this method, the BHs are approximately in a stationary state in their
own comoving frame and hence a large amount of spurious gravitational
waves are not included, in contrast to the initial data prepared by
the simple moving-puncture approach.

\begin{figure}[t]
%\vspace{-8mm}
\epsfxsize=3.in
\leavevmode
\epsffile{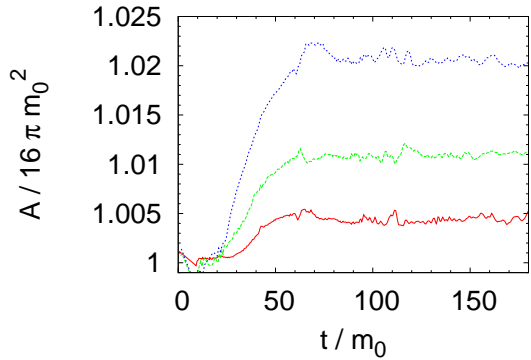}
%\end{center}
\vspace{-5mm}
\caption{Area of apparent horizon as a function of time before
  collision for $v=0.9$ and $bv/(m_0\Gamma)=4.708$ with different grid
  resolution, $h/m_0=0.0625$ (dotted curve), 0.050 (dashed curve), and
  0.042 (solid curve).
\label{FIG1.5}}
\end{figure}

With this initial data, apparent horizons are located approximately
for $r_a=m_0/2$ at $t=0$. The area of each horizon is $\approx 16\pi
m_0^2$ within the error of $O(10^{-3})$ for $x_a \agt 100m_0$ (thus
the bare mass of the BHs would be $\approx m_0$, and hence, a fraction
of kinetic energy in the total mass of the system is $1-\Gamma^{-1}$).
Furthermore, the area of the apparent horizon remains approximately
$16\pi m_0^2$ during evolution before collision
(cf. Fig. \ref{FIG1.5}). Therefore, it is reasonable to expect that
numerical solution obtained by the simulation provides an approximate
solution within an error of $\sim 1\%$.  In the future work, we will
perform simulations using improved initial data which satisfies the
constraints, obtained by computing correction terms, $\phi$ and
$\delta K_{ij}$, to strictly validate the present strategy.

%\section{Numerical results}

\noindent
{\bf\em III Numerical results}: 
For numerical simulation, we use {\small SACRA} code recently developed by our
group \cite{SACRA}. In {\small SACRA}, the Einstein equations are solved in a
modified version of BSSN (Baumgarte-Shapiro-Shibata-Nakamura)
formalism \cite{BSSN} with a fourth-order finite differencing scheme
in space and time and with an adaptive mesh refinement algorithm (at
refinement boundaries, second-order interpolation scheme is partly
adopted and hence the convergence may reduce to be second order).  The
moving-puncture approach is adopted for following moving BHs
\cite{BHBH}. Gravitational waves are computed by extracting the
outgoing part of the Newman-Penrose quantity (the so-called $\Psi_4$).
Properties of the BHs such as mass and spin are determined by
analyzing area and circumferential radii of apparent horizons.  The
details of our schemes, formulation, gauge conditions, and methods for
the analysis are described in \cite{SACRA}. This reference also shows
that {\small SACRA} can successfully simulate merger of two equal-mass
BHs. Because an adaptive mesh refinement algorithm is implemented, the
moving BHs can be computed accurately by preparing a high-resolution
domain appropriately around the BHs; in the present work, we prepare
10 refinement levels.  Indeed, we performed test simulations in which
a single high-velocity BH is boosted with $v=0.8$ and 0.9, and found
that our code can follow such high-velocity BH for more than $500
m_0$; e.g., we checked that the area of apparent horizon converges at
approximately fourth order with improving grid resolution and the 
error is within $\sim 0.1\%$ and 1\% for $v=0.8$ and 0.9,
respectively, for the grid resolution with $h=m_0/20$
(e.g., Fig. \ref{FIG1.5}).

\begin{figure}[t]
\vspace{-8mm}
\epsfxsize=3.5in
\leavevmode
\epsffile{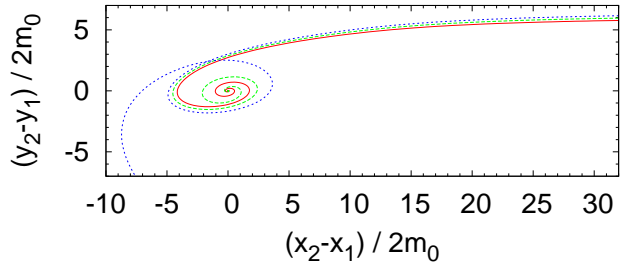}
%\end{center}
\vspace{-15mm}
\caption{Trajectories of relative position of moving punctures for
  $v=0.9$ and $b/m_0=6.0$, 6.2, and 6.4 (solid, dashed, and dotted
  curves).  ($bv/m_0\Gamma=4.708$, 4.865, and 5.021.)
\label{FIG1}}
\end{figure}

Numerical simulations are performed for $v=0.6$, 0.7, 0.8, and 0.9 and
$x_0/m_0=160$ changing the impact parameter $b$.  As expected from
Eq.~(\ref{spin}), two BHs should merge after collision for $b < 4
m_0\Gamma/v$, and hence, the critical value of the impact parameter
for the BH formation (hereafter $b_{\rm crit}$) should be close to $4
m_0\Gamma/v$. For this reason, the value of $b$ is chosen in the
range, $4 \alt bv/ (m_0\Gamma) \alt 5.5$, with the basic step size of
$b$ being $0.1m_0$ (near $b=b_{\rm crit}$, the step size of $0.05m_0$ is
partly used).

Numerical results depend only weakly on the initial separation for
given values of $v$ and $b$, and for a given grid resolution.  Indeed,
we performed detailed test simulations for $v=0.8$ and 0.9, and
$x_0/m_0=80$, 128, and 160. If $x_0/m_0$ is changed from 160 to 128,
the value of $b_{\rm crit}$, which is one of the most important
outputs in this work, decreases only by $\alt 0.05m_0$, and from
$x_0/m_0=160$ to 80, by $\alt 0.1m_0$. Recall that the larger value of
the initial separation results in the smaller initial constraint
violation. Thus, the weak dependence of the numerical results on $x_0$
indicates that the initial constraint violation only weakly affects
the numerical results.

The numerical simulations are performed for different grid resolutions
as $h/m_0=0.075$, 0.0625, 0.050, and 0.042. The outer boundaries along
each axis are located at $L \approx 770 m_0$ for all the grid
resolution. The value of $L/c$ is longer than the duration that the
simulation is carried out, and hence, spurious effects from the outer
boundaries are excluded.  We find that the value of $b_{\rm crit}$
depends only weakly on the grid resolution (it increases only by $\sim
0.1m_0$ ($\sim 1.5\%$) if we change $h/m_0$ from 0.0625 to 0.042).
The area of a BH formed after the merger, and total energy and angular
momentum dissipated by gravitational waves ($\Delta E$ and $\Delta J$)
depend more strongly on the grid resolution.  However, our results
indicate a convergence slightly better than second order with
improving the grid resolution as long as $h/m_0 \leq 0.0625$ and $0.6
\leq v \leq 0.9$, although the convergence becomes slow for $v \agt
0.9$. This is natural because the coordinate radius of the apparent
horizon in the $x$-axis direction for each BH is proportional to
$\Gamma^{-1}$. Obviously, the better grid resolution is necessary for
the ultra-high velocity collision $v \rightarrow 1$, but this is
beyond scope of this paper.

\begin{figure}[t]
\epsfxsize=3.in
\leavevmode
\epsffile{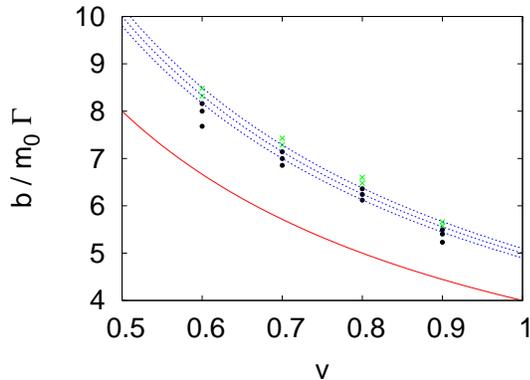}
%\end{center}
\vspace{-2mm}
\caption{Summary of the final outcomes after the collision of two BHs
  in the parameter space of $(v, b)$.  The circles denote that a BH is
  formed after collision, whereas the crosses do not. The solid curve
  denotes $b/(m_0\Gamma/v)=4.0$, and the dashed curves are 4.9, 5.0,
  and 5.1.
\label{FIG2}}
\end{figure}

Figure~\ref{FIG1} plots trajectories of moving punctures for $v=0.9$,
$x_0=160m_0$, and $bv/(m_0 \Gamma)=4.708$, 4.865, and 5.021. Here, we
plot relative position $[(x_2-x_1)/2m_0, (y_2-y_1)/2m_0]$ on the
orbital plane. For the first two impact parameters, the BH is formed
after the collision, whereas for the other, each BH escapes from the
center after scattering. For the small values of $b$, two BHs form a
bound orbit when the orbital separation becomes small enough.  For a
sufficiently small value of $b$, such as $bv/(m_0 \Gamma)\alt 4.7$,
two BHs merge within one orbit. For a value of $b$ close to $b_{\rm
  crit}$, two BHs rotate around each other for more than one orbits
before two BHs merge, as shown in Fig.~\ref{FIG1}. For $b> b_{\rm
  crit}$, two BHs rotate around each other for small
separation. However, they do not constitute a bound orbit because of
the large centrifugal force, and eventually, each BH escapes from the
center.

To summarize the outcomes after the collision, we generate
Fig.~\ref{FIG2} which shows a parameter space composed of $(v, b)$.
We plot the circles for the case that two BHs merge after the
collision, whereas the crosses are plotted when two BHs do not
merge. The solid curve denotes $b = 4m_0\Gamma/v$ for which the
nondimensional spin parameter of the system is unity at $t=0$. The
three dashed curves denote $bv/m_0\Gamma=4.9$, 5.0, and 5.1.  Figure
\ref{FIG2} clarifies that for $b\alt b_{\rm crit} \approx (5.0 \pm
0.1) m_0 \Gamma/v$, a BH is formed after the collision for the chosen
velocity, $0.6 \leq v \leq 0.9$. Extrapolating the result for $v
\rightarrow 1$ under the assumption that the discovered relation holds
even for $v \rightarrow 1$, the maximum impact parameter is determined
to be $(2.50\pm 0.05) M_0$.

\begin{figure}[t]
\epsfxsize=3.in \leavevmode \epsffile{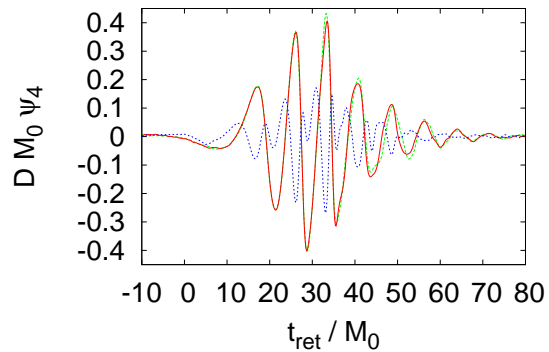}
\vspace{-5mm}
%\caption{Outgoing part of the Newman-Penrose quantity for $v=0.9$ and
%  $bv/m_0\Gamma=4.708$.  The solid and dashed curves denote $l=m=2$ and
%  $l=m=4$ modes, respectively. $t_{\rm ret}=0$ approximately
%  corresponds to the time of the onset of merger. $D$ denotes a 
%distance between the source and observer. 
\caption{Outgoing part of the Newman-Penrose quantity for $v=0.8$ and
  $bv/(m_0\Gamma)=5.088$.  The solid and dotted curves denote $l=m=2$
  and $l=m=4$ modes for the best-resolved run, and the dashed curve
  denotes $l=m=2$ modes for the second-best run. $t_{\rm ret}=0$
  approximately corresponds to the time of the onset of merger. $D$
  denotes a distance between the source and observer.
\label{FIG3}}
\end{figure}

Figure \ref{FIG3} plots outgoing part of the Newman-Penrose quantity
for $v=0.8$ and $bv/m_0\Gamma=5.088$.  $l=m=2$ and $l=m=4$ modes are
plotted for the best-resolved run with $h=0.042m_0$. For $l=m=2$, we
also show the result for $h=0.05m_0$, which agrees with the
best-resolved result with a small error.  Note that the waveforms are
qualitatively similar irrespective of the value of $v$ as far as $b
\approx b_{\rm crit}$, although the maximum amplitude steeply
increases when $b$ approaches $b_{\rm crit}$.  Figure \ref{FIG3} shows
that gravitational waves are efficiently emitted after the onset of
collision: When two BHs approach each other, amplitude of
gravitational waves gradually increases. As the separation of two BHs
becomes sufficiently small, they constitute a bound orbit and
quasiperiodic gravitational waves are emitted. After substantial
fraction of gravitational waves are emitted, two BHs merge to be a new
BH, and then, ring-down gravitational waves associated with
fundamental quasinormal modes are emitted for $t_{\rm ret} \agt
40M_0$. In this case, the formed BH is rapidly rotating with the spin
parameter $\approx 0.73 \pm 0.02$, and hence, the damping time scale
is longer than that for nonrotating BHs \cite{L85}. Because the
orbital velocity is very large, higher-multipole components of
gravitational waves are also enhanced significantly (cf. the waveform
for $l=m=4$).

Total energy and angular momentum dissipated by gravitational waves
are $\approx 25 \pm 5\%$ and $\approx 65 \pm 5\%$ of the initial
energy and angular momentum, respectively, for $b \alt b_{\rm crit}$
and for $v=0.9$. The totally emitted gravitational radiation for $b
\sim b_{\rm crit}$ slightly decreases with decreasing $v$, but still,
$\Delta E/M_0 \agt 20\%$ and $\Delta J/J \agt 60\%$ for $0.6 \leq v
\leq 0.8$.  $l=|m|=4$ modes contribute to $\Delta E$ and $\Delta J$ by
$\approx 10$--15\% and by $\approx 15$--20\%, respectively, for $b
\sim b_{\rm crit}$.  It should be noted that in the limit $b
\rightarrow b_{\rm crit}$, total amount of gravitational radiation may
be slightly larger than that presented here, because the lifetime of
the formed binary orbit could be longer.  We note that the results for
the total amount of gravitational radiation are consistent with the
mass and spin of the BH finally formed within an acceptable error for
the best-resolved runs: The mass and angular momentum of the BHs
estimated from apparent horizon are always smaller than those expected
from gravitational radiation (i.e., $M_0-\Delta E$ and $J-\Delta J$)
by $\alt 0.05M_0$ and $\alt 0.1J$, respectively, for the best-resolved
run. The error is larger for the larger value of $v$.  The reason for
this error is that energy and angular momentum are dissipated
spuriously by numerical effects associated with finite grid resolution.
However, our results show a behavior of convergence slightly better
than second order with improving grid resolution.

%\section{Discussion and summary}

\noindent
{\bf\em IV Discussion and summary}:
We find that the largest value of the impact parameter for the BH formation
after the collision is $b_{\rm crit}\approx (2.50 \pm 0.05) M_0/v$. 
For such value of the impact parameter, the initial value of 
the spin parameter of the system is 
\beqn
{J \over M_0^2} =1.25 \pm 0.03.
\eeqn
For the BH formation, the spin parameter of the formed BH should be smaller 
than unity if the cosmic censorship holds \cite{Wald}. 
This implies that a large fraction of the angular momentum 
is dissipated by gravitational radiation during the collision. 
We estimate the total amount of the angular momentum and energy 
dissipated by gravitational radiation are $\Delta J=(0.65 \pm 0.05)J$ and 
$\Delta E=(0.25 \pm 0.05)M_0$, respectively. The expected spin parameter 
of the formed BH is 
\beqn
\approx {J-\Delta J \over (M_0-\Delta E)^2} \approx (0.6\pm 0.1)
{J \over M_0^2}. 
\eeqn
Thus, even if the spin parameter of the system is initially 
1.25, the resulting value at BH formation is smaller than unity. 
As this discussion clarifies, gravitational radiation increases 
the impact factor by $\sim 25\%$, and as a result, the cross section 
by $\sim 50\%$. It is also worth noting that the 
formed BH does not have an extremely large spin $\sim 1$, but
approximately $0.8 \pm 0.1$ even for $b \alt b_{\rm crit}$. 

In this work, we adopt initial data which satisfies the constraint
equations only approximately. Although the violation is tiny
(cf. Fig. \ref{FIG0}), this error produces a small error in estimation
of the critical cross section, and $\Delta E$ and $\Delta J$ of
gravitational waves. To determine these quantities strictly, it is
necessary to perform simulations using improved initial condition.

This work is a first step toward detailed understanding of 
high-velocity collision of two BHs in higher-dimensional spacetime. We
plan to develop a numerical code for higher-dimensional spacetimes. 

%\begin{acknowledgments}

\noindent
{\bf\em Acknowledgments}: We thank T. Shiromizu and K. Maeda for
helpful discussions and comments.  This work was in part supported by
Monbukagakusho Grant No. 19540263.

%\end{acknowledgments}

\end{document}